\documentstyle[12pt]{article}

\begin{document}
\parskip 1ex
\setcounter{page}{1}
\oddsidemargin 0pt
\evensidemargin 0pt
\topmargin -40pt
%
\newcommand{\be}{\begin{equation}}
\newcommand{\ee}{\end{equation}}
\newcommand{\beq}{\begin{eqnarray}}
\newcommand{\eeq}{\end{eqnarray}}
\def\a{\alpha}
\def\b{\beta}
\def\g{\gamma}
\def\G{\Gamma}
\def\d{\delta}
\def\e{\epsilon}
\def\z{\zeta}
\def\h{\eta}
\def\th{\theta}
\def\k{\kappa}
\def\l{\lambda}
\def\L{\Lambda}
\def\m{\mu}
\def\n{\nu}
\def\x{\xi}
\def\X{\Xi}
\def\p{\pi}
\def\P{\Pi}
\def\r{\rho}
\def\s{\sigma}
\def\S{\Sigma}
\def\t{\tau}
\def\f{\phi}
\def\F{\Phi}
\def\c{\chi}
\def\w{\omega}
\def\W{\Omega}
\def\de{\partial}

\def\pct#1{(see Fig. #1.)}

\begin{titlepage}
\hbox{\hskip 12cm ROM2F-98/18  \hfil}
\hbox{\hskip 12cm hep-th/9806129 \hfil}
\vskip 1.4cm
\begin{center}  {\Large  \bf   Consistent \ and \ Covariant  \ Anomalies  \vskip
.6cm  in \ Six-dimensional \ Supergravity}

\vspace{1.8cm}
 
{\large \large  Fabio Riccioni\footnote{I.N.F.N. Fellow} \ and \ Augusto
Sagnotti}
\vspace{0.6cm}

{\sl Dipartimento di Fisica, \ \ Universit{\`a} di Roma \ ``Tor Vergata'' \\
I.N.F.N.\ - \ Sezione di Roma \ ``Tor Vergata'', \\ Via della Ricerca
Scientifica , 1 \ \ \ 00133 \ Roma \ \ ITALY}
\end{center}
\vskip 1.5cm

\abstract{In this note we clarify some issues in six-dimensional $(1,0)$
supergravity coupled to vector and tensor multiplets. In particular, 
we show that, while the
low-energy equations embody tensor-vector couplings that contribute only to
gauge anomalies, the divergence of
the energy-momentum tensor is properly non-vanishing.  In addition, we show
how to revert to a supersymmetric formulation in terms of covariant
non-integrable field equations that embody corresponding covariant anomalies.}

\vskip 2.5cm
\begin{center}
{( June , \ 1998 )}
\end{center}
\vfill
\end{titlepage}
\makeatletter
\@addtoreset{equation}{section}
\makeatother
\renewcommand{\theequation}{\thesection.\arabic{equation}}
\addtolength{\baselineskip}{0.3\baselineskip} 

\section{Introduction}

Six-dimensional $(1,0)$ supergravity has attracted a large interest in recent
years for a number of reasons \cite{review}.  On the one hand, vector multiplets
coupled to variable numbers of tensor multiplets arise naturally in perturbative
type-I vacua \cite{bs}, and therefore, via duality, play a ubiquitous role
in non-perturbative  string phenomena.  On the other hand, the field equations
have revealed the explicit realization of a peculiar aspect of the physics of
branes. Namely, branes wrapped on vanishing cycles in the internal manifold may
result in the exotic phenomenon of transitions related to  tensionless
strings \cite{witten}, 
and indeed some peculiar singularities in the gauge couplings  of
$(1,0)$ models in moduli space \cite{as,ns2} can be ascribed 
to phase transitions
\cite{dmw} whereby a string becomes tensionless 
\cite{tensionless}.  On a more technical side, these equations present the novel
feature of a Green-Schwarz mechanism implemented by terms present in the
low-energy effective action, at least for the  gauge part of the residual
anomaly.  One is thus facing a case of unprecedented complexity in supergravity
constructions, whereby the model is determined by Wess-Zumino  conditions
\cite{wz},  rather than by the usual requirement of local supersymmetry. 
Moreover, as pointed out in \cite{frs}, the algebra contains a two-cocycle and 
the resulting equations are not unique. By and large, this is a remarkable
laboratory for current algebra, where one can play explicitly with anomalous
symmetries and their consequences.  

The present note is devoted to some aspects of current algebra related to the
energy-momentum tensor that, although rather simple, are somewhat surprising and
were not noticed in \cite{frs}. The corresponding analysis is carried out in the
next Section.  An additional, related problem  has to do with the formulation of
the resulting equations, that were originally derived in \cite{as} to lowest
order in the fermion couplings by requiring local supersymmetry. The subsequent
work of \cite{fms} and
\cite{frs} has developed the consistent formulation, but one can actually revert
to a covariant formulation, at the price of having non-integrable
field equations.  The relation between the two sets of equations is one more
instance of the link between covariant and consistent anomalies in field
theory
\cite{bz}.  Once more, here  the anomalies are  induced by local couplings of the
two-forms, and everything is totally explicit.   The resulting covariant
equations turn into one another under local  supersymmetry and complete to all
orders in the fermi fields the results of
\cite{as}.

\section{The energy-momentum tensor of six-dimensional $(1,0)$ supergravity}

In six-dimensional $(1,0)$ supergravity coupled to an arbitrary number of tensor
and vector multiplets, the $n$ scalars in the tensor  multiplets parametrize the
coset space $SO(1,n)/SO(n)$, and are described by the $SO(1,n)$ matrix
\cite{romans}
\be V =\pmatrix{v_r \cr x^m_r}\quad .
\ee All spinors are symplectic Majorana-Weyl. In particular, the tensorinos
$\chi^m$  are right-handed, while the gravitino $\Psi_\m$ and the gauginos $\l$
are left-handed. The tensor fields $B^r_{\m\n}$ are valued in the fundamental 
representation of $SO(1,n)$ and satisfy the (anti)self-duality conditions
\be G_{rs} H^{s \m\n\r} =\frac{1}{6e} \e^{\m\n\r\a\b\g} H_{r \a\b\g}\quad ,
\label{selfdual}
\ee where $G_{rs} =v_r v_s + x^m_r x^m_s$. 
Moreover, their field strengths include Chern-Simons
3-forms for the vector fields according to
\be H^r  = d B^r -c^{rz} \w_z \quad ,
\ee where the $c^{rz}$ are constants that determine the gauge part of the
residual anomaly polynomial
\be {\cal A} = - \sum_{rs} \eta_{rs}\  c^{rx} \  c^{sy} \ 
tr_x ( F \wedge F ) \  tr_y ( F
\wedge F )
\ee and $z$ runs over the various factors of the gauge group \cite{as}.  Gauge
invariance of $H^r$ then requires that 
\be
\delta B^r = c^{rz} tr_z (\L dA) \quad .
\ee 
The complete field equations were determined in \cite{frs} from the
commutator of two supersymmetry transformations on the fermi fields, in
the spirit of \cite{schwarz,romans}. The
resulting model, however, has gauge and supersymmetry  anomalies 
(to be canceled by fermion loops) related by 
Wess-Zumino consistency conditions and is not unique. 
Aside from subtleties related to the
(anti)self-dual antisymmetric tensors \cite{pst},  
all field equations may be derived from
\beq  
e^{-1}{\cal{L}} & & =-\frac{1}{4}R +\frac{1}{12}G_{rs} H^{r \m\n\r}
H^s_{\m\n\r} -\frac{1}{4} \de_\m v^r \de^\m v_r-\frac{1}{2} v_r c^{rz} tr_z
(F_{\m\n} F^{\m\n}) 
\nonumber\\ & & -\frac{1}{8e}
\e^{\m\n\a\b\g\delta} c_r^z B^r_{\m\n} tr_z (F_{\a\b} F_{\g\delta}) \nonumber \\
& &  -\frac{i}{2}\bar{\Psi}_\m \g^{\m\n\r} D_\n [\frac{1}{2}(\w +\hat{\w} )]
\Psi_\r -\frac{i}{8}v_r [H+\hat{H}]^{r \m\n\r}(\bar{\Psi}_\m \g_\n \Psi_\r)
\nonumber \\ & & +\frac{i}{48} v_r [H+\hat{H} ]^r_{\a\b\g} (\bar{\Psi}_\m
\g^{\m\n\a\b\g}\Psi_\n )+\frac{i}{2} \bar{\chi}^m \g^\m D_\m (\hat{\w})
\chi^m \nonumber \\ & & -\frac{i}{24}v_r \hat{H}^r_{\m\n\r} (\bar{\chi}^m
\g^{\m\n\r}
\chi^m ) +\frac{1}{4}x^m_r [\de_\n v^r +\hat{\de_\n v^r} ](\bar{\Psi}_\m \g^\n
\g^\m \chi^m) \nonumber\\ & & -\frac{1}{8} x^m_r [H+\hat{H}]^{r \m\n\r} (
\bar{\Psi}_\m
\g_{\n\r}
\chi^m )+\frac{1}{24}x^m_r [H+\hat{H}]^{r \m\n\r} (\bar{\Psi}^\a \g_{\a\m\n\r}
\chi^m ) \nonumber\\ 
& & +\frac{1}{8}(\bar{\chi}^m \g^{\m\n\r} \chi^m )(\bar{\Psi}_\m
\g_\n \Psi_\r )-\frac{1}{8}(\bar{\chi}^m \g^\m \chi^n )(\bar{\chi}^m \g_\m
\chi^n ) \nonumber\\ 
& & +
\frac{i}{2\sqrt{2}}  v_r c^{rz} tr_z [(F+\hat{F})_{\n\r}(\bar{\Psi}_\m
\g^{\n\r}\g^\m \l )] 
 +\frac{1}{\sqrt{2}}x^m_r c^{rz} tr_z [(\bar{\chi}^m \g^{\m\n}
\l )\hat{F}_{\m\n} ] \nonumber\\  
& & +iv_r c^{rz} tr_z (\bar{\l} \g^\m
\hat{D}_\m \l )+\frac{i}{12} x^m_r x^m_s \hat{H}^r_{\m\n\r} c^{sz} tr_z
(\bar{\l}\g^{\m\n\r} \l )
\nonumber\\ & &  +\frac{1}{16}v_r c^{rz}tr_z (\bar{\l} \g_{\m\n\r} \l
)(\bar{\chi}^m
\g^{\m\n\r} \chi^m ) \nonumber\\  & & - \frac{i}{8}(\bar{\chi}^m
\g_{\m\n}\Psi_\r )x^m_r c^{rz} tr_z (\bar{\l} \g^{\m\n\r} \l ) 
 - \frac{i}{2} x^m_r c^{rz} tr_z [(\bar{\chi}^m \g^\m \g^\n \l ) (\bar{\Psi}_\m
\g_\n \l )]
\nonumber\\ & & -\frac{1}{8} v_r c^{rz} tr_z [(\bar{\chi}^m \l )
(\bar{\chi}^m \l)]- \frac{3}{16}v_r c^{rz}tr_z [(\bar{\chi}^m \g_{\m\n} \l )
(\bar{\chi}^m
\g^{\m\n}
\l )] 
\nonumber\\ & & -\frac{3}{4}\frac{x^m_r c^{rz}x^n_s c^{sz}}{v_t c^{tz}} tr_z
[(\bar{\chi}^m \l )(\bar{\chi}^n \l )] 
 +\frac{1}{8}\frac{x^m_r c^{rz} x^n_s c^{sz}}{v_t c^{tz}} tr_z [(\bar{\chi}^m
\g_{\m\n}
\l )(\bar{\chi}^n \g^{\m\n}\l )]
\nonumber\\ & & +\frac{1}{4} (\bar{\Psi}_\m \g_\n \Psi_\r )(\bar{\l} \g^{\m\n\r}
\l )  -\frac{1}{2} v_r v_s c^{rz}c^{s z^\prime} tr_{z,z^\prime} [(\bar{\l}\g_\m
\l^\prime )(\bar{\l} \g^\m \l^\prime ) ]\nonumber\\
& & + \frac{\a}{2}c^{rz} c_r^{z^\prime} tr_{z,z^\prime} [(\bar{\l}
\g_\m
\l^\prime )(\bar{\l} \g^\m \l^\prime )]
\quad,
\label{completelag}
\eeq 
where $\a$ is an arbitrary parameter whose role was discussed at length 
in \cite{frs}.
The variation of ${\cal L}$ with respect to the supersymmetry
transformations
\beq & & \delta e_\m{}^a =-i(\bar{\e} \g^a \Psi_\m ) \quad,\nonumber\\  & &
\delta B^r_{\m\n} =i v^r (\bar{\Psi}_{[\m} \g_{\n]} \e )+ \frac{1}{2} x^{mr}
(\bar{\chi}^m
\g_{\m\n} \e )-2c^{rz} tr_z (A_{[\m}\delta A_{\n]}) \quad,
\nonumber\\  & & \delta v_r = x^m_r (\bar{\chi}^m \e )\quad,\nonumber\\  & &
\delta A_\m = -\frac{i}{\sqrt{2}} (\bar{\e} \g_\m \l ) \quad ,\nonumber\\  & &
\delta \Psi_\m =\hat{D}_\m \e +\frac{1}{4} v_r \hat{H}^r_{\m\n\r}
\g^{\n\r}\e -\frac{3i}{8} \g_\m \chi^n (\bar{\e} \chi^n ) -\frac{i}{8} \g^\n
\chi^n (\bar{\e} \g_{\m\n} \chi^n )+\frac{i}{16} \g_{\m\n\r} \chi^n (\bar{\e} 
\g^{\n\r} \chi^n ) \nonumber \\ & & \quad \quad - \frac{9i}{8} v_r c^{rz} tr_z
[\l (\bar{\e} \g_\m \l)] +  
\frac{i}{8} v_r c^{rz} tr_z [\g_{\m\n} \l (\bar{\e} \g^\n \l )] - \frac{i}{16}
v_r c^{rz} tr_z [\g^{\n\r} \l (\bar{\e}
\g_{\m\n\r} \l )] \quad ,\nonumber\\  
& & \delta \chi^m =
\frac{i}{2} x^m_r (\hat{\de_\m v^r} ) \g^\m \e +
\frac{i}{12} x^m_r \hat{H}^r_{\m\n\r} \g^{\m\n\r}\e +
\frac{1}{2} x^m_r c^{rz} tr_z [ \g_\m \l (\bar{\e} \g^\m \l ) ] \quad
,\nonumber\\  & & \delta \l =-\frac{1}{2\sqrt{2}}\hat{F}_{\m\n} \g^{\m\n} \e  -
\frac{1}{2} \frac{x^m_r c^{rz}}{v_s c^{sz}} (\bar{\chi}^m \l ) \e 
 - \frac{1}{4} \frac{x^m_r c^{rz}}{v_s c^{sz}} (\bar{\chi}^m \e ) \l  
\nonumber \\ & & \quad \quad + \frac{1}{8} \frac{x^m_r c^{rz}}{v_s c^{sz}}
(\bar{\chi}^m \g_{\m\n} \e ) \g^{\m\n}
\l 
\eeq  gives the supersymmetry anomaly
\beq  {\cal{A}}_\e & &=c_r^z c^{r z^\prime} tr_{z, z^\prime} \lbrace -\frac{1}{4}
\e^{\m\n\a\b\g\delta}\delta_\e A_\m A_\n F^\prime_{\a\b} F^\prime_{\g\delta}
-\frac{1}{6}
\e^{\m\n\a\b\g\delta} \delta_\e A_\m F_{\n\a} 
\w^\prime_{\b\g\delta} \nonumber\\ & & +\frac{i e}{2} \delta_\e A_\m F_{\n\r}
(\bar{\l}^\prime
\g^{\m\n\r} 
\l^\prime )+\frac{i e}{2} \delta_\e A_\m (\bar{\l} \g^{\m\n\r} \l^\prime )
F^\prime_{\n\r}  + ie\delta_\e A_\m (\bar{\l}\g_\n \l^\prime ) F^{\prime \m\n}
\nonumber\\ & & +
\frac{e}{32} \delta_\e e_\m{}^a (\bar{\l} \g^{\m\n\r} \l )(\bar{\l}^\prime
\g_{a \n\r} \l^\prime )  -\frac{e}{2\sqrt{2}} \delta_\e A_\m (\bar{\l} \g^\m
\g^\n \g^\r 
\l^\prime )(\bar{\l}^\prime \g_\n \Psi_\r )  \nonumber\\  & & + \frac{e x^m_s
c^{s z^\prime}}{v_t c^{t z^\prime}} [-\frac{3 i}{2\sqrt{2}}
 \delta_\e A_\m (\bar{\l} \g^\m \l^\prime )(\bar{\l}^\prime \chi^m )
 -\frac{i}{4 \sqrt{2}} \delta_\e A_\m (\bar{\l} \g^{\m\n\r} \l^\prime
)(\bar{\l}^\prime
\g_{\n\r}\chi^m ) \nonumber\\ & & - \frac{i}{2\sqrt{2}} \delta_\e A_\m 
(\bar{\l} \g_\n
\l^\prime )(\bar{\l}^\prime \g^{\m\n} \chi^m )] +\delta e_\m{}^a 
\frac{\a e e^\m{}_a}{2} 
(\bar{\l}\g^\n \l^\prime )(\bar{\l}\g_\n \l^\prime )\nonumber\\
& &+e \delta A_\m [-i\a
\hat{F}_{\n\r}(\bar{\l}^\prime \g^{\m\n\r}\l^\prime )+   i\a(\bar{\l}
\g^{\m\n\r}\l^\prime ) 
\hat{F}^\prime_{\n\r} - 6i\a(\bar{\l}\g^\n 
\l^\prime ) \hat{F}^\prime_{\m\n}]\nonumber\\ 
& & + \delta A_\m
\frac{e x^m_s c^{s z^\prime}}{v_t c^{t z^\prime}} [-i \a \sqrt{2}
(\bar{\l}\g^\m \l^\prime ) (\bar{\l}^\prime \chi^m )
+\frac{i\a}{2\sqrt{2}}(\bar{\l}\g_{\n\r}\chi^m )(\bar{\l}^\prime 
\g^{\m\n\r}\l^\prime )]\nonumber\\ 
& & +\delta A_\m \frac{e x^m_s c^{s
z}}{v_t c^{t z}}  [\frac{i \a}{\sqrt{2}}(\bar{\l}\g^\m \l^\prime
)(\bar{\l}^\prime
\chi^m ) -\frac{i \a}{2\sqrt{2}}(\bar{\l} \g^{\m\n\r} \l^\prime )(\bar{\l}^\prime
\g_{\n\r} \chi^m )\nonumber\\ & & +\frac{i \a}{\sqrt{2}}(\bar{\l}\g_\n \l^\prime
)(\bar{\l}^\prime \g^{\m\n}
\chi^m ) ]
\rbrace \quad,
\label{theanomaly}
\eeq while the variation of ${\cal L}$ with respect to vector gauge
transformations gives  the consistent gauge anomaly
\be  {\cal{A}}_\L =- \frac{1}{4} \e^{\m\n\a\b\g\delta} c_r^z c^{rz^\prime} tr_z
(\L
\de_\m A_\n ) tr_{z^\prime} (F_{\a\b} F_{\g\delta} )\quad .
\label{consanomaly}
\ee 

The commutator of two supersymmetry transformations on the  fermi fields closes
on the equations of (\ref{completelag}), generating all local symmetry
transformations, as well as the extra two-cocycle
\beq
 \delta_{extra(\a)} \l & &\equiv [\delta_1 , \delta_2 ]_{extra(\a)} \l =
\frac{c_r^z c^{r z^\prime}}{v_s c^{sz}} tr_{z^\prime} [-\frac{1}{4}(\bar{\e}_1
\g_\m \l^\prime )(\bar{\e}_2
\g_\n \l^\prime ) \g^{\m\n} \l \nonumber\\  & &-\frac{\a}{2} (\bar{\l} \g_\m
\l^\prime )(\bar{\e}_1 \g_\n \l^\prime )
\g^{\m\n} \e_2  +\frac{\a}{16}(\bar{\l}\g_{\m\n\r}\l^\prime )(\bar{\e}_1 \g^\r 
\l^\prime ) \g^{\m\n} \e_2 \nonumber\\  & &+\frac{\a}{16} (\bar{\l} \g_\r
\l^\prime )(\bar{\e}_1
\g^{\m\n\r} \l^\prime ) \g_{\m\n} \e_2   + \frac{1-\a}{4} (\bar{\l} \g_\m
\l^\prime ) (\bar{\e}_1
\g^\m \l^\prime ) \e_2 -(1 \leftrightarrow 2)  
\nonumber\\ & & + \frac{1-\a}{16} (\bar{\e}_1 \g^\m \e_2 )(\bar{\l}^\prime
\g_{\m\n\r} \l^\prime  )\g^{\n\r}\l ] \label{centralcharge}
\eeq for the gauginos.  The presence of the arbitrary parameter $\alpha$ reflects the
freedom of adding to the anomaly the variation of a local functional, consistently
with all Wess-Zumino conditions. In six dimensions these
close only on the field equations of the gaugini,
and the two-cocyle grants the consistency of the
construction for all values of $\alpha$ \cite{frs}.
It would be interesting to perform a cohomological 
analysis in superspace of this system along the lines of \cite{bb}.

The gauge anomaly ${\cal A}_\L = \delta_\L {\cal L}$ naturally satisfies the
condition
\be {\cal A}_\L= -tr (\L D_\m J^\m )\quad, \label{div1}
\ee where $J^\m =0$ is the complete field equation of the vector field. One can
similarly show that the supersymmetry anomaly is related to the field equation
of the gravitino, that we write succinctly
${\cal J}^\m =0$, according to
\be {\cal A}_\e = -(\bar{\e} D_\m {\cal J}^\m ) \quad .\label{div2}
\ee

We would like to stress that the Noether identities (\ref{div1}) and
(\ref{div2}) relate the anomalies to the equations of the fields whose
transformations contain derivatives. This observation has 
a natural application to
gravitational anomalies, that we would now like to elucidate. In fact, 
in analogy with the previous cases one would expect that
\be {\cal A}_{\xi} = \delta_{\xi}{\cal L}= 2 \xi_\m D_\n T^{\m\n} \quad ,
\ee where the variation of the metric under general 
coordinate transformations is
\be
\delta g_{\m\n} = -\xi^\a \de_\a g_{\m\n} -g_{\a\n} \de_\m \xi^\a -g_{\m\a} 
\de_\n \xi^\a \quad .
\ee 
Thus, for models without gravitational anomalies one
would expect that the divergence of the energy-momentum tensor  vanish. Actually,
this is no longer true if other anomalies are present, since  all
fields, not only the metric, have derivative variations under  coordinate
transformations. For instance, in a theory with gauge and
supersymmetry anomalies, the gravitational anomaly is actually
\be 
{\cal A}_{\xi} = \delta_{\xi}{\cal L}= 2 \xi_\n D_\m T^{\m\n} +
\xi_\n tr (A^\n D_\m J^\m ) + \xi_\n  (\bar{\Psi}^\n D_\m
{\cal J}^\m )\quad. \label{emtnoether}
\ee 
In particular, in our case we are not accounting for gravitational
anomalies, that would  result in higher-derivative couplings, and indeed one can
verify that the divergence of the energy-momentum tensor does not vanish, but
satisfies the relation
\be
D_\m T^{\m\n} =-\frac{1}{2} tr(A^\n D_\m J^\m )- \frac{1}{2} 
(\bar{\Psi}^\n D_\m {\cal J}^\m )\quad. \label{emtdiv}
\ee

\section{Covariant field equations and covariant anomalies} 

It is well known
that consistent and covariant gauge anomalies are related  by the divergence of
a local functional \cite{bz}. In six dimensions the residual covariant gauge
anomaly is \cite{as}
\be {\cal A}_\L^{cov}=\frac{1}{2} \e^{\m\n\a\b\g\delta} c^{rz}c_r^{z^\prime}tr_z
(\L F_{\m\n}) tr_{z^\prime} (F^\prime_{\a\b} F^\prime_{\g\delta} ) 
\quad , \label{covanomaly}
\ee and is related to the consistent anomaly by a local counterterm,
\be {\cal A}_\L^{cons} +tr [\L D_\m f^\m ] ={\cal A}_\L^{cov}\quad
,\label{covcons}
\ee where
\be f^\m =c_r^z c^{r z^\prime}  \lbrace -\frac{1}{4}
\e^{\m\n\a\b\g\delta} A_\n tr_{z^\prime}( F^\prime_{\a\b} F^\prime_{\g\delta} )
-\frac{1}{6} \e^{\m\n\a\b\g\delta}  F_{\n\a} \ 
\w^\prime_{\b\g\delta} \rbrace \quad .\label{fmu}
\ee  Comparing eq. (\ref{fmu}) with eq. (\ref{theanomaly}) one can see that, to
lowest order in the fermi fields,
\be {\cal A}_\e = tr(\delta_\e A_\m f^\m )  \quad,
\ee and this implies that the transition from consistent  to  covariant anomalies
turns a model with a supersymmetry anomaly into another without any
\cite{as,fms}. Indeed, six-dimensional supergravity coupled to vector and tensor
multiplets was originally formulated in this fashion in \cite{as} to lowest
order in the fermi fields, extending the results of Romans 
\cite{romans}\footnote{The complete coupling to a single tensor multiplet, as well as to 
vector and hyper multiplets, was originally constructed in \cite{ns1} for the
special case of vanishing residual anomaly.}. The
resulting vector equation is not integrable. Moreover, the corresponding gauge
anomaly is not the gauge variation of a local functional and does not satisfy
Wess-Zumino consistency conditions.

This result can be generalized naturally, if somewhat tediously, to include terms
of all
orders in the fermi fields. The complete supersymmetry anomaly of eq.
(\ref{theanomaly}) has the form
\be {\cal A}_\e = tr(\delta_\e A_\m f^\m ) +\delta_\e e_\m{}^a g^\m{}_a \quad ,
\ee where  to lowest order $f^\m$ is defined in eq. (\ref{fmu}).  Modifying the
vector equation so that
\be 
(eq. \ A^\m )_{(cov)} \equiv J^\m_{(cov)}
=\frac{\delta {\cal L}}{\delta A_\m } -f^\m \quad ,
\label{modifiedvector}
\ee 
and similarly for the Einstein equation, the resulting theory is
supersymmetric but no longer integrable. The covariant vector field equation
is
\beq & & 2 D_\n (v_r F^{\m\n} )-2 G_{rs}\hat{H}^{s\m\n\r}F_{\n\r}
-\frac{i}{2}v_r (\bar{\Psi}_\a \g^{\a\b\m\n\r}\Psi_\b )F_{\n\r}\nonumber\\ & &
+\frac{i}{2}v_r (\bar{\chi}^m \g^{\m\n\r} \chi^m ) F_{\n\r} -x^m_r
(\bar{\Psi}_\a \g^{\a\m\n\r} \chi^m ) F_{\n\r}-i x^m_r x^m_s c^{s z^\prime} 
tr_{z^\prime}
(\bar{\l}^\prime \g^{\m\n\r} \l^\prime )  
F_{\n\r}\nonumber\\ & & +i \sqrt{2} D_\n [v_r
(\bar{\Psi}_\r \g^{\m\n} \g^\r \l )]+
\sqrt{2} D_\n [x^m_r (\bar{\chi}^m \g^{\m\n} \l )]\nonumber\\ & &
-\frac{i}{2} F_{\n\r}c_r^{z^\prime} tr_{z^\prime} (\bar{\l}^\prime
\g^{\m\n\r} 
\l^\prime ) -\frac{i}{2} 
c_r^{z^\prime}tr_{z^\prime}[ (\bar{\l} \g^{\m\n\r} \l^\prime )
F^\prime_{\n\r}]  - i c_r^{z^\prime}[ (\bar{\l}\g_\n \l^\prime )  F^{\prime
\m\n}]
\nonumber\\  & &  +\frac{1}{2\sqrt{2}} c_r^{z^\prime}tr_{z^\prime} [(\bar{\l}
\g^\m \g^\n \g^\r 
\l^\prime )(\bar{\l}^\prime \g_\n \Psi_\r )]  \nonumber\\  & & + \frac{x^m_s
c^{s z^\prime}}{v_t c^{t z^\prime}} c_r^{z^\prime} tr_{z^\prime} [\frac{3
i}{2\sqrt{2}} (\bar{\l} \g^\m \l^\prime )(\bar{\l}^\prime \chi^m )
 +\frac{i}{4 \sqrt{2}} (\bar{\l} \g^{\m\n\r} \l^\prime )(\bar{\l}^\prime
\g_{\n\r}\chi^m ) \nonumber\\ & & +\frac{i}{2\sqrt{2}} (\bar{\l} \g_\n
\l^\prime )(\bar{\l}^\prime \g^{\m\n} \chi^m )] \nonumber \\ & & +
c_r^{z^\prime} tr_{z^\prime} [i\a \hat{F}_{\n\r}(\bar{\l}^\prime
\g^{\m\n\r}\l^\prime )-  i\a(\bar{\l} \g^{\m\n\r}\l^\prime ) 
\hat{F}^\prime_{\n\r} + 6i\a(\bar{\l}\g^\n 
\l^\prime ) \hat{F}^\prime_{\m\n}]\nonumber\\ & & + c_{r}^{z^\prime} \frac{x^m_s
c^{s z^\prime}}{v_t c^{t z^\prime}} tr_{z^\prime}[i \a \sqrt{2} (\bar{\l}\g^\m
\l^\prime ) (\bar{\l}^\prime \chi^m )
-\frac{i\a}{2\sqrt{2}}(\bar{\l}\g_{\n\r}\chi^m )(\bar{\l}^\prime 
\g^{\m\n\r}\l^\prime )]\nonumber\\ & & + c_{r}^{z^\prime} \frac{x^m_s c^{s
z}}{v_t c^{t z}} tr_{z^\prime} [-\frac{i \a}{\sqrt{2}}(\bar{\l}\g^\m \l^\prime
)(\bar{\l}^\prime
\chi^m ) +\frac{i \a}{2\sqrt{2}}(\bar{\l} \g^{\m\n\r} \l^\prime )(\bar{\l}^\prime
\g_{\n\r} \chi^m )\nonumber\\ & & -\frac{i \a}{\sqrt{2}}(\bar{\l}\g_\n \l^\prime
)(\bar{\l}^\prime \g^{\m\n}
\chi^m ) ] = 0 \quad ,\label{covvectoralpha}
\eeq
and completes the results in \cite{as} to all orders in the fermi fields.  Its
divergence satisfies
\be tr (\L D_\m J^\m_{(cov)} )=-{\cal A}_\L^{cov}\quad ,
\ee where ${\cal A}_\L^{cov}$ contains higher-order fermi terms:
\beq  
{\cal{A}}_\L^{cov} & &=c^{rz}c_r^{z^\prime}tr_{z,z^\prime}\lbrace
\frac{1}{2} \e^{\m\n\a\b\g\delta} (\L F_{\m\n})  (F^\prime_{\a\b}
F^\prime_{\g\delta} ) \nonumber\\  
& & +i e \L F_{\n\r} (\bar{\l}^\prime
\g^{\m\n\r}D_\m 
\l^\prime )+\frac{i e}{2} \L D_\m (\bar{\l} \g^{\m\n\r} \l^\prime )
F^\prime_{\n\r}  + ie \L D_\m [ (\bar{\l}\g_\n \l^\prime ) F^{\prime \m\n}]
\nonumber\\  
& &  -\frac{e}{2\sqrt{2}} \L D_\m [ (\bar{\l} \g^\m \g^\n
\g^\r 
\l^\prime )(\bar{\l}^\prime \g_\n \Psi_\r )  ]\nonumber\\  
& & + e \L D_\m \lbrace
\frac{ x^m_s c^{s z^\prime}}{v_t c^{t z^\prime}} [-\frac{3 i}{2\sqrt{2}}
(\bar{\l} \g^\m \l^\prime )(\bar{\l}^\prime \chi^m )
 -\frac{i}{4 \sqrt{2}} (\bar{\l} \g^{\m\n\r} \l^\prime )(\bar{\l}^\prime
\g_{\n\r}\chi^m ) \nonumber\\ & & - \frac{i}{2\sqrt{2}}  (\bar{\l} \g_\n
\l^\prime )(\bar{\l}^\prime \g^{\m\n} \chi^m )] \rbrace \nonumber\\
& & + e \L D_\m [-i\a
\hat{F}_{\n\r}(\bar{\l}^\prime \g^{\m\n\r}\l^\prime )+   i\a(\bar{\l}
\g^{\m\n\r}\l^\prime ) 
\hat{F}^\prime_{\n\r} - 6i\a(\bar{\l}\g^\n 
\l^\prime ) \hat{F}^\prime_{\m\n}]\nonumber\\ & & 
+ e \L D_\m \lbrace
\frac{x^m_s c^{s z^\prime}}{v_t c^{t z^\prime}} [-i \a \sqrt{2}
(\bar{\l}\g^\m \l^\prime ) (\bar{\l}^\prime \chi^m )
+\frac{i\a}{2\sqrt{2}}(\bar{\l}\g_{\n\r}\chi^m )(\bar{\l}^\prime 
\g^{\m\n\r}\l^\prime )]\rbrace \nonumber\\ 
& & +e \L D_\m \lbrace \frac{x^m_s c^{s
z}}{v_t c^{t z}}  [\frac{i \a}{\sqrt{2}}(\bar{\l}\g^\m \l^\prime
)(\bar{\l}^\prime
\chi^m ) -\frac{i \a}{2\sqrt{2}}(\bar{\l} \g^{\m\n\r} \l^\prime )(\bar{\l}^\prime
\g_{\n\r} \chi^m )\nonumber\\ & & +\frac{i \a}{\sqrt{2}}(\bar{\l}\g_\n \l^\prime
)(\bar{\l}^\prime \g^{\m\n}
\chi^m ) ]\rbrace
\rbrace \quad .
\label{completecovanomaly}
\eeq 

Finally, one can study the divergence of the Rarita-Schwinger and Einstein 
equations in the covariant model. To this end, let us begin by stating that
the derivation of Noether identities for a system of non-integrable 
equations does
not present difficulties of principle, since these involve only first 
variations. 
Indeed, the only difference with respect to
the standard case of integrable equations is that now 
$\delta {\cal L}$ is not an exact differential in field space.  
Still, all invariance principles reflect themselves
in linear dependencies of the field equations.  Thus, for instance, 
with the covariant
equations obtained from the consistent ones by the redefinition of eq. 
(\ref{modifiedvector}) and by
\be 
(eq. \ e^\m{}_a )_{(cov)}=\frac{\delta {\cal L}}{\delta e_\m{}^a } -g^\m{}_a 
\quad ,\ee
the total $\delta_\e {\cal L}$ vanishes by construction. 
The usual procedure then proves that the 
divergence of the Rarita-Schwinger
equation vanishes for any value of the parameter $\a$. 
On the other hand, the divergence
of the energy-momentum tensor presents some subtleties that we 
would now like to 
describe.  In particular, it vanishes to lowest order in the fermi couplings, 
while it gives a covariant non-vanishing result if all fermion 
couplings are taken into
account.  The subtlety has to do with the transformation of the vector 
under general coordinate transformations,
\be
\delta _\xi A_\m = - \xi^\a \de_\a A_\m - \de_\m \xi^\a A_\a \quad ,
\ee
and with the corresponding full (off-shell) form of the identity of 
eq. (\ref{emtnoether}).
Starting again from the consistent equations, one finds
\be {\cal A}_{\xi} = \delta_{\xi}{\cal L}= 2 \xi_\n D_\m T^{\m\n} +
\xi_\n tr (A^\n D_\m J^\m ) + \xi_\n tr (F^{\m\n} J_\m ) 
+ \xi_\n  (\bar{\Psi}^\n D_\m
{\cal J}^\m )\quad. \label{emtnoetheroff}
\ee 
Reverting to the covariant form eliminates the divergence of the Rarita-Schwinger equation
and alters the vector equation, so that the third
term has to be retained. The final result is then
\be
D_\m T^{\m\n}_{(cov)} = -\frac{1}{2}  tr(A^\n D_\m J^\m_{(cov)} )-
\frac{1}{2}tr(f_\m  F^{\m\n} ) -\frac{1}{2}tr( A^\n D_\m f^\m )
-\frac{1}{2} e^{\n a} D_\m g^\m{}_a \quad ,
\ee
and is nicely verified by our equations.  In particular, this implies that, to lowest
order in the fermi couplings, the divergence of $T^{\m\n}_{(cov)}$ vanishes.

\end{document}